\documentclass[aps,preprint,nofootinbib,superscriptaddress,prc]{revtex4}
\usepackage{amssymb}

\usepackage{epsfig}
\usepackage[bookmarksnumbered,bookmarksopen,colorlinks,citecolor=blue,linkcolor=blue] {hyperref}
\begin{document}


\title{Improved Kelson-Garvey mass relations for proton-rich nuclei}

\author{Junlong Tian}
 \email{tianjunlong@gmail.com}
 \affiliation{ School of Physics and Electrical Engineering,
Anyang Normal University, Anyang 455000, People's Republic of
China }
\author{Ning Wang}
\affiliation{ Department of Physics, Guangxi Normal University,
Guilin 541004, People's Republic of China }
\author{Cheng Li}
\affiliation{  School of Physics and Electrical Engineering,
Anyang Normal University, Anyang 455000, People's Republic of
China }
\affiliation{ Department of Physics, Guangxi Normal University,
Guilin 541004, People's Republic of China }
\author{Jingjing Li}
\affiliation{  School of Physics and Electrical Engineering,
Anyang Normal University, Anyang 455000, People's Republic of
China }

\begin{abstract}

The improved Kelson-Garvey (ImKG) mass relations are proposed from
the mass differences of mirror nuclei. The masses of 31 measured
proton-rich nuclei with $7\leq A\leq41$ and $-5\leq (N-Z)\leq-3$ can
be remarkably well reproduced by using the proposed relations, with
a root-mean-square deviation of 0.398 MeV, which is much smaller
than the results of Kelson-Garvey (0.502 MeV) and Isobar-Mirror mass
relations (0.647 MeV). This is because many more masses of
participating nuclei are involved in the ImKG mass relations for
predicting the masses of unknown proton-rich nuclei. The masses for
144 unknown proton-rich nuclei with $6\leq A\leq74$ are predicted by
using the ImKG mass relations. The one- and two-proton separation
energies for these proton-rich nuclei and the diproton emission are
investigated simultaneously.

\end{abstract}

\maketitle

\section{\label{sec:level1}INTRODUCTION}

Nuclear mass (or binding energy), which reflects directly the sum
effect of the strong, weak and electromagnetic interactions among
the participating nucleons, plays a vital role in nuclear physics
and astrophysics. In nuclear physics, nuclear masses are of great
importance for the development of nuclear structure models and the
prediction of exotic decay modes and new symmetries. With the help
of radioactive beam facilities, a variety of nuclei near or beyond
the proton drip line have been produced, including the heaviest and
most proton-rich doubly magic nucleus $^{48}$Ni \cite{Dossat07}. In
astrophysics, the investigation of nucleosynthesis and energy
generation in the rapid proton capture (rp)-process
\cite{Wallace81,Wormer94} and $\nu$p process \cite{Frohlich06}
depends on the masses of proton-rich nuclei. The rp-process path
approaches the proton drip line for heavy nuclei. The
proton-separation energy $S_{p}$ is of particular interest to the
$\nu$p process as it is a process which could resolve the
long-standing underproduction of light p-nuclei. Accurate
predictions for the masses of proton drip-line nuclei is therefore
urgently required for understanding the rp-process and
$\nu$p-process \cite{Bsun11,Schatz98}.

Available nuclear mass formulas include some global and local
formulas. Some global nuclear mass models have been successfully
established, such as the Duflo-Zuker model \cite{DZ95}, the finite
range droplet model (FRDM) \cite{Moller95}, the Skyrme
Hartree-Fock-Bogoliubov theory  \cite{Gori01,Gori10,Geng05}, the
Weizs$\ddot{a}$cker-Skyrme (WS) mass model
\cite{Ning10,liumin11,Ning11}. For drip line nuclei, the differences
between the calculated masses from these different models are quite
large. The local mass formulas are generally based on algebraic or
systematic approaches. They predict the masses of unknown nuclei
from the masses of known neighboring nuclei, such as the
Audi-Wapstra systematics \cite{Audi93,Audi95,Audi03}, the
Garvey-Kelson mass relations
\cite{Garv66,Janec74,Mona77,Janec85,Bare05,Bare08,Hirs08,Mora11},
and the mass relations based on the residual proton-neutron
interactions \cite{Fugj10,Jiangh10,Fugj11,Jiangh12}. The main
difficulty of the available local mass formulas is that the model
errors rapidly increase for nuclei far from the measured nuclei. On
the other hand, the concept of symmetry in physics is a very
powerful tool for understanding the behavior of nature. The isospin
symmetry discovered by Heisenberg plays an important role in nuclear
physics. In the absence of Coulomb interactions between the protons,
a perfectly charge-symmetric and charge-independent nuclear force
would result in the binding energies of mirror nuclei being
identical. It is therefore necessary to establish some new mass
relations for description of the masses of proton-drip line nuclei
according to the mirror effect coming from the isospin symmetry in
nuclear physics.

In this work, we attempt to establish some new mass relations for
the description of the mass of proton-rich nuclei. The paper is
organized as follows: In Sec. II, the improved Kelson-Garvey (ImKG)
mass relation are proposed from the mass differences of mirror
nuclei including the higher-order mirror nuclei and ordinary mirror
nuclei near the N=Z line. In Sec. III, the masses of 144 proton-rich nuclei are calculated with the proposed method. The one-
and two-proton separation energies of nuclei, and the diproton
emission of nuclei are investigated simultaneously. Finally, a
summary is given in Sec. IV.

\section{\label{sec:level2} Improved Kelson-Garvey mass relations}

\subsection{\label{sec:level1} Kelson-Garvey (KG) mass relations }

A simple and successful procedure has been introduced by Kelson and
Garvey \cite{KG66,KG88}. It is a mass relation which connects masses
of higher-order mirror nuclei with ordinary mirror nuclei near the
$N=Z$ line,
\begin{eqnarray}
 B(A,-Y)-B(A,Y)= \sum_{j=-(Y-1)}^{+(Y-1)}\left[ B(A+j,Y=-1)-B(A+j,Y=1) \right ],
 \end{eqnarray}
where $B(A,Y)$ is the binding energy of a nucleus which can be
expressed as a function of mass number $A$ and isotopic number
$Y=N-Z$, and $j$ is to be increased by a step of 2.
\begin{figure}
\includegraphics[angle=-0,width= 0.5\textwidth]{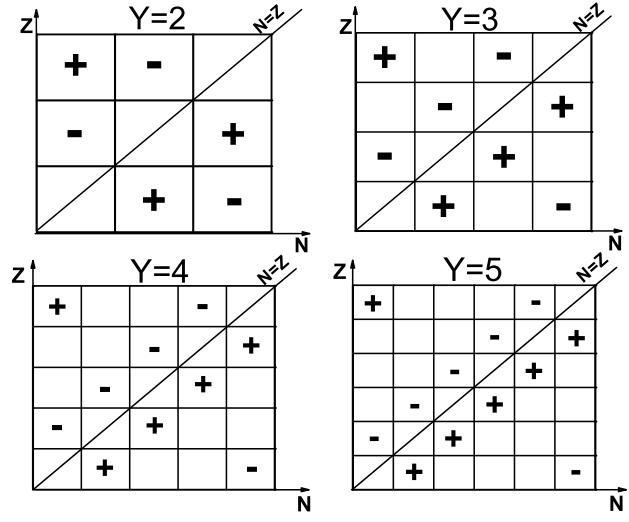}
 \caption{ Schematic representation of the charge-symmetric Kelson-Garvey mass relation Eq.(1) for four examples, $Y=2-5$.
The boxes represent nuclei from the nuclidic chart, and the plus and
minus signs indicate that the respective mass values have to be
added or subtracted.}
\end{figure}

The binding energy differences between the most proton-rich and most
neutron-rich members of an isospin multiplet can hence be estimated
from known binding energy differences between $Y=1$ mirror nuclei. A
heuristic proof of this relation can again be obtained from an
independent particle picture with fourfold degenerate Hartree-Fock
or Nilsson-like single-particle levels. Fig. 1 shows the schematic
representation of the charge-symmetric KG mass relation Eq.(1) for
four examples with $Y=2-5$. The boxes represent nuclei from the
nuclidic chart, and the plus and minus signs indicate that the
respective mass values have to be added or subtracted. The binding
energy differences for the $Y = 1$ mirror nuclei up to $A = 75$ were
taken from the experimental mass table given in Ref. \cite{AME2011}.

\subsection{\label{sec:level1}Improved Kelson-Garvey (ImKG) mass relations}

According to the KG mass relation of Eq.(1), the binding energy
difference between a pair of mirror nuclei can be written as
\begin{eqnarray}
\nonumber B(A,-2)-B(A,2)\approx B(A-1,-1)-B(A-1,1)\\
+B(A+1,-1)-B(A+1,1),
\end{eqnarray}
for $Y=2$ case, and \begin{eqnarray}
\nonumber B(A,-3)-B(A,3)&&\approx B(A-2,-1)-B(A-2,1)\\
\nonumber&&+B(A,-1)-B(A,1)\\
&&+B(A+2,-1)-B(A+2,1),
\end{eqnarray}
for $Y=3$ case.

Applying Eq.(2), we rewrite Eq.(3) as follows (see Fig.2)
\begin{eqnarray}
\nonumber B(A,-3)-B(A,3)\approx B(A-1,-2)-B(A-1,2)\\
+B(A+2,-1)-B(A+2,1),
\end{eqnarray}
or
\begin{eqnarray}
\nonumber B(A,-3)-B(A,3)\approx B(A-2,-1)-B(A-2,1)\\
+B(A+1,-2)-B(A+1,2).
\end{eqnarray}

\begin{figure}
\includegraphics[angle=-0,width= 0.9\textwidth]{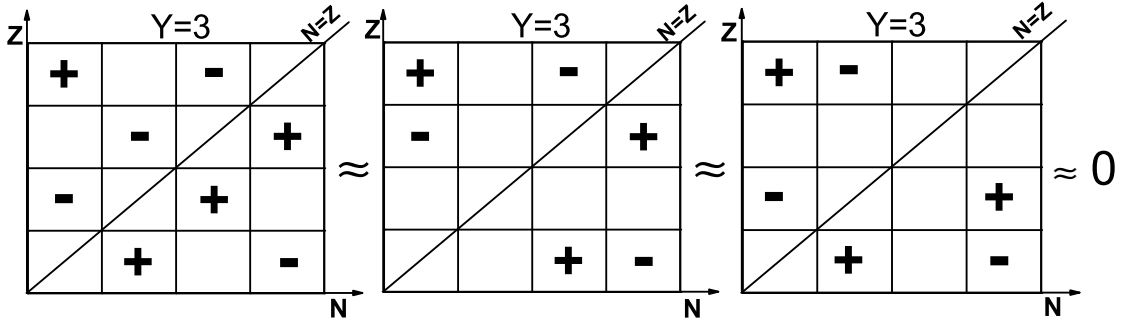}
 \caption{Schematic representation of the ImKG relation Eq.(3, 4, 5) for $Y=3$ cases.}
\end{figure}

Through the summation of Eq.(3), Eq.(4) and Eq.(5), we can obtain the
following expression as an improved Kelson-Garvey mass relation (see
$Y=3$ case in Fig.3)
\begin{eqnarray}
 \nonumber 3[B(A,-3)-B(A,3)]&&\approx 2B(A-2,-1)-2B(A-2,1)\\
\nonumber&&+B(A,-1)-B(A,1)\\
\nonumber&&+2B(A+2,-1)-2B(A+2,1)\\
\nonumber&&+B(A-1,-2)-B(A-1,2)\\
&&+B(A+1,-2)-B(A+1,2),
\end{eqnarray}
or
\begin{eqnarray}
\nonumber && 3B(A,-3)-B(A+1,-2)-2B(A+2,-1)+0B(A+3,0)\\
\nonumber&&-B(A-1,-2)-B(A,-1)+0B(A+1,0)+2B(A+2,1)\\
\nonumber&&-2B(A-2,-1)+0B(A-1,0)+B(A,1)+B(A+1,2)\\
&&-0B(A-3,0)+2B(A-2,1)+B(A-1,2)-3B(A,3)\approx 0.
\end{eqnarray}

In the same fashion, one can predict the binding energy of unknown
proton-rich nucleus $B(A,-4)$ by expression (see $Y=4$ case in
Fig.3)
\begin{eqnarray}
 \nonumber 7[B(A,-4)-B(A,4)]&&\approx 4[B(A-3,-1)-B(A-3,1)]\\
\nonumber&&+2[B(A-1,-1)-B(A-1,1)]\\
\nonumber&&+2[B(A+1,-1)-B(A+1,1)]\\
\nonumber&&+4[B(A+3,-1)-B(A+3,1)]\\
\nonumber&&+2[B(A-2,-2)-B(A-2,2)]\\
\nonumber&&+[B(A,-2)-B(A,2)]\\
\nonumber&&+2[B(A+2,-2)-B(A+2,2)]\\
\nonumber&&+[B(A-1,-3)-B(A-1,3)]\\
&&+[B(A+1,-3)-B(A+1,3)].
\end{eqnarray}

Following a format similar to Eqs.(6) and (8), the following general
relationship is obtained as the improved Kelson-Garvey mass relation
\begin{eqnarray}
\nonumber (2^{Y-1}-1)[B(A,-Y)-B(A,Y)] && \approx \sum_{i=1}^{(Y-1)} \{ 2^{Y-i-1}[B(A-(Y-i),-i)-B(A-(Y-i),i)]\\
\nonumber && +2^{Y-i-2}\sum_{j=-(Y-i-2)}^{(Y-i-2)}[B(A+j,-i)-B(A+j,i)]\\
&&+2^{Y-i-1}[B(A+(Y-i),-i)-B(A+(Y-i),i)] \},
\end{eqnarray}
where $Y=2, 3, 4, ....$; $i=1, 2, 3, ..., (Y-1)$, and $(Y-i-2)\geq0$
otherwise the term
$2^{Y-i-2}\sum_{j=-(Y-i-2)}^{(Y-i-2)}[B(A+j,-i)-B(A+j,i)]$ is
canceled; $j=-(Y-i-2),-(Y-i-2)+2, ..., (Y-i-2)$ is to be increased by step of 2.
\begin{figure}
\includegraphics[angle=-0,width= 0.5\textwidth]{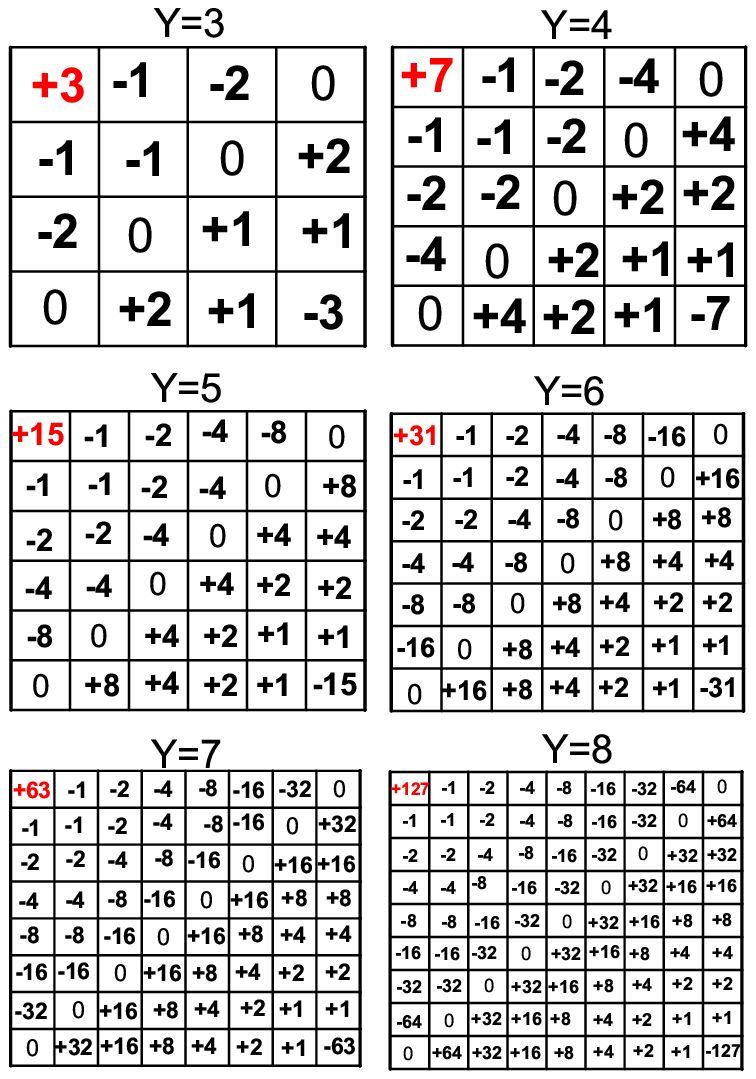}
 \caption{(Color online) Schematic representation of the improved Kelson-Garvey mass relations given by Eq.(9) for 6 examples with $Y=3-8$.}
\end{figure}

\subsection{\label{sec:level1}  Isobar-Mirror mass relations  }

Isobar-mirror mass relations are deduced based on three assumptions
as follows: 1) The difference between the binding energies of mirror
nuclei is only due to the Coulomb interaction. It is known that in
the absence of Coulomb interactions between the protons, a perfectly
charge-symmetric and charge-independent nuclear force would result
in the binding energies of mirror nuclei being identical
\cite{Ormand97,Lenzi09,shlomo78}; 2) The Coulomb energy difference
between a pair of mirror nuclei is proportional to $Y$, the same as
the assumption used in Ref. \cite{Ormand97},
\begin{eqnarray}
\Delta B  = E_{C}(A,-Y)-E_{C}(A,Y) = b(A,Y,i)Y,
\end{eqnarray}
in which the label $i$ represents the microscopic corrections
connected to the nuclear force, which are neglected; 3) The
coefficient of proportionality which indeed depends on $A$, may be
considered as independent of $Y$ for a given $A$. In Refs.
\cite{Ormand89,Ormand97}, it was found that the $b$ coefficients are
roughly constant (see the Tables 3-7 in Ref. \cite{Ormand89}) for a
given $A$, and Ormand obtained an empirical formula $b =
0.710A^{2/3}-0.946$ MeV by fitting to 116 experimental data with an
rms deviation of 102 keV \cite{Ormand97}.

In the following calculation by using isobar-mirror mass relations,
one does not need to know the exact value of $b$ coefficients. The
binding energies difference for a pair odd-$A$ mirror nuclei with
different $Y=1, 3, 5,...$ reads
\begin{eqnarray}
\Delta B(Y=1)=B(A,1)-B(A,-1)=b,
\end{eqnarray}
\begin{eqnarray}
\Delta B(Y=3)=B(A,3)-B(A,-3)=3b,
\end{eqnarray}
\begin{eqnarray}
\Delta B(Y=5)=B(A,5)-B(A,-5)=5b,
\end{eqnarray}
\begin{eqnarray}
\nonumber ......
\end{eqnarray}

One sees that
\begin{eqnarray}
\frac{\Delta B(Y=1)}{\Delta
B(Y=3)}=\frac{B(A,1)-B(A,-1)}{B(A,3)-B(A,-3)}=\frac{1}{3}.
\end{eqnarray}
If the binding energies for nuclei $B(A,3)$, $B(A,-1)$ and $B(A,1)$
are available, the binding energy for unmeasured proton-rich nucleus
$B(A,-3)$ can be predicted by using
\begin{eqnarray}
B(A,-3)=B(A,3)+3[B(A,-1)-B(A,1)].
\end{eqnarray}
Similarly, one can predict the binding energy of unknown proton-rich
nucleus $B(A,-5)$ by expressions
\begin{eqnarray}
B(A,-5)=B(A,5)+5[B(A,-1)-B(A,1)],
\end{eqnarray}
and
\begin{eqnarray}
B(A,-5)=B(A,5)+\frac{5}{3}[B(A,-3)-B(A,3)].
\end{eqnarray}
Taking the mean value for $B(A,-5)$, we can obtain the binding
energy of $B(A,-5)$
\begin{eqnarray}
B(A,-5)=B(A,5)+\frac{5}{2}[B(A,-1)-B(A,1)]+\frac{5}{6}[B(A,-3)-B(A,3)].
\end{eqnarray}

Based on the isobar-mirror mass relations, the binding energy of a
proton-rich odd-$A$ nucleus can be described by
\begin{eqnarray}
 B(A,-Y)=B(A,Y)+\sum_{t }\frac{2Y}{t(Y-1) } \left
 [B(A,-t)-B(A,t) \right ]
\end{eqnarray}
with $Y=3,5,7,9,...$ and $t=1,3,5,...,Y-2$. Similarly, the binding
energy for a proton-rich even-$A$ nucleus is expressed as
\begin{eqnarray}
 B(A,-Y)=B(A,Y)+\sum_{t} \frac{2 Y}{t(Y-2)}[B(A,-t)-B(A,t)]
\end{eqnarray}
with $Y=4, 6, 8, 10, ...$ and $t=2,4,6,...,Y-2$.

\begin{figure}
\includegraphics[angle=-0,width= 0.7\textwidth]{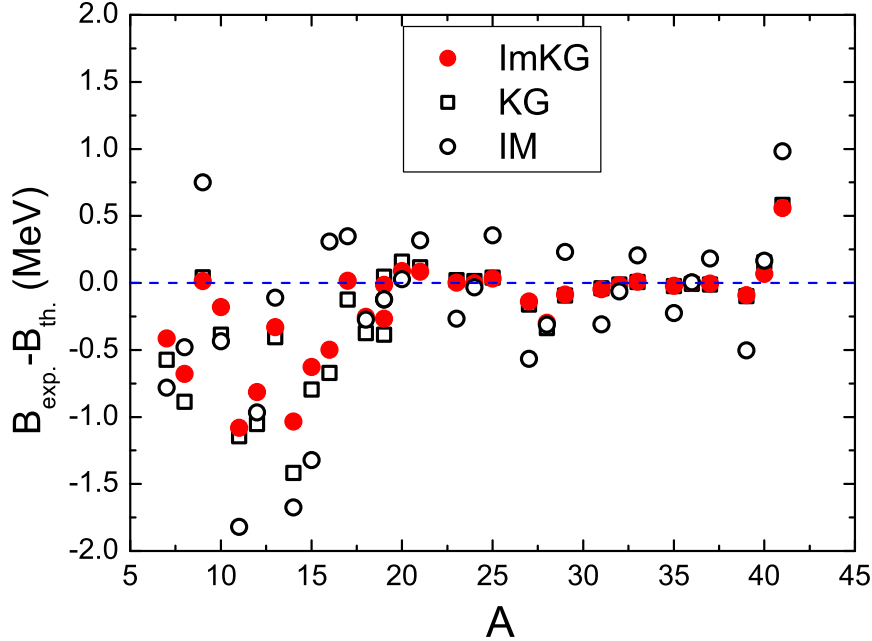}
 \caption{(Color online) Difference between the experimental binding energies and predictions for 31 measured proton-rich nuclei with ImKG approach (solid circles), KG \cite{KG88} (open squares) and IM (open circles), respectively.}
\end{figure}

\section{\label{sec:level2} Results and discussion}
\subsection{\label{sec:level2} Test of the ImKG mass relations }
We study the binding energies of 92 pairs measured mirror nuclei
based on tabulated masses (AME2011) \cite{AME2011}, including 35
($Y=1$), 26 ($Y=2$), 18 ($Y=3$), 12 ($Y=4$) and 1 ($Y=5$) pairs of
mirror nuclei. Fig. 4 shows the differences between the experimental
binding energies and predicted absolute binding energies with three
different theoretical models for 31 experimental measured
proton-rich nuclei with $7\leq A \leq41$ and $-5\leq Y \leq-3$. The
experimental binding energies are taken from the mass table AME2011
\cite{AME2011}. The solid circles are the results of the ImKG
approach. The open squares and open circles denote the results of KG
\cite{KG88} and IM, respectively. From Fig.4, one sees that the
deviation between the ImKG calculation and the experimental data is
the smallest one. We also note that for two nuclei $^{11}$N and
$^{14}$F, the deviations are larger than one MeV from all three
methods.

The root-mean-square (rms) deviations between the experimental binding energies \cite{AME2011} and model predictions
\begin{eqnarray}
  \sigma(M)=\left [\frac{1}{m}\sum \left (B_{\rm exp}^{(i)}-B_{\rm
th}^{(i)}\right )^2 \right ] ^{1/2}
\end{eqnarray}
are calculated. The obtained rms deviation with respect to these 31
known masses are 0.398 MeV with ImKG mass relation, which is much
smaller than the results from KG (0.502 MeV) \cite{KG88} and IM
(0.647 MeV) mass relations. There are $Y(Y+1)-1$ and $2Y+1$
participating nuclei involved in the ImKG predictions and in the KG
mass relation, respectively. Because many more participating nuclei
are involved in the ImKG predictions, the reliability for predicting
the binding energies of unknown proton-rich nuclei is significantly
improved.

In Table I, we list the calculated binding energies and mass
excesses by using the ImKG mass relations for 144 unknown
proton-rich nuclei with  $5\leq A\leq 74$. The estimated
uncertainties are given in parentheses in keV. The uncertainty of
the predicted binding energy is determined by the uncertainty of the
experimental binding energies of the participating nuclei. The
isotopic number $Y=N-Z$ is listed in column 2. Columns 5 and 6 list
the neutron-rich analog and the measured mass excess with its
associated error in parentheses in keV.

\subsection{\label{sec:level2} One- and two-proton separation energies of unknown proton-rich nuclei}
Based on the binding energies predicted by the ImKG mass relations,
we evaluate one-proton and two-proton separation energies for
proton-rich nuclei:
\begin{eqnarray}
S_{p}(A, Z)=B(A,Z)-B(A-1,Z-1),
\end{eqnarray}
\begin{eqnarray}
S_{2p}(A, Z)=B(A,Z)-B(A-2,Z-2).
\end{eqnarray}
The predicted separation energies for one-proton (S$_{p}$) and
two-proton (S$_{2p}$) with the ImKG mass relations for 144
proton-rich nuclei with $5 \leq A \leq 74$ are given in columns 7
and 8 in Table I, respectively. In the calculation, we adopt the
experimental data if the mass of the nucleus is available. The
masses of unknown nuclei are predicted by using the ImKG mass
relations.

\begin{figure}
\includegraphics[angle=-0,width= 0.7\textwidth]{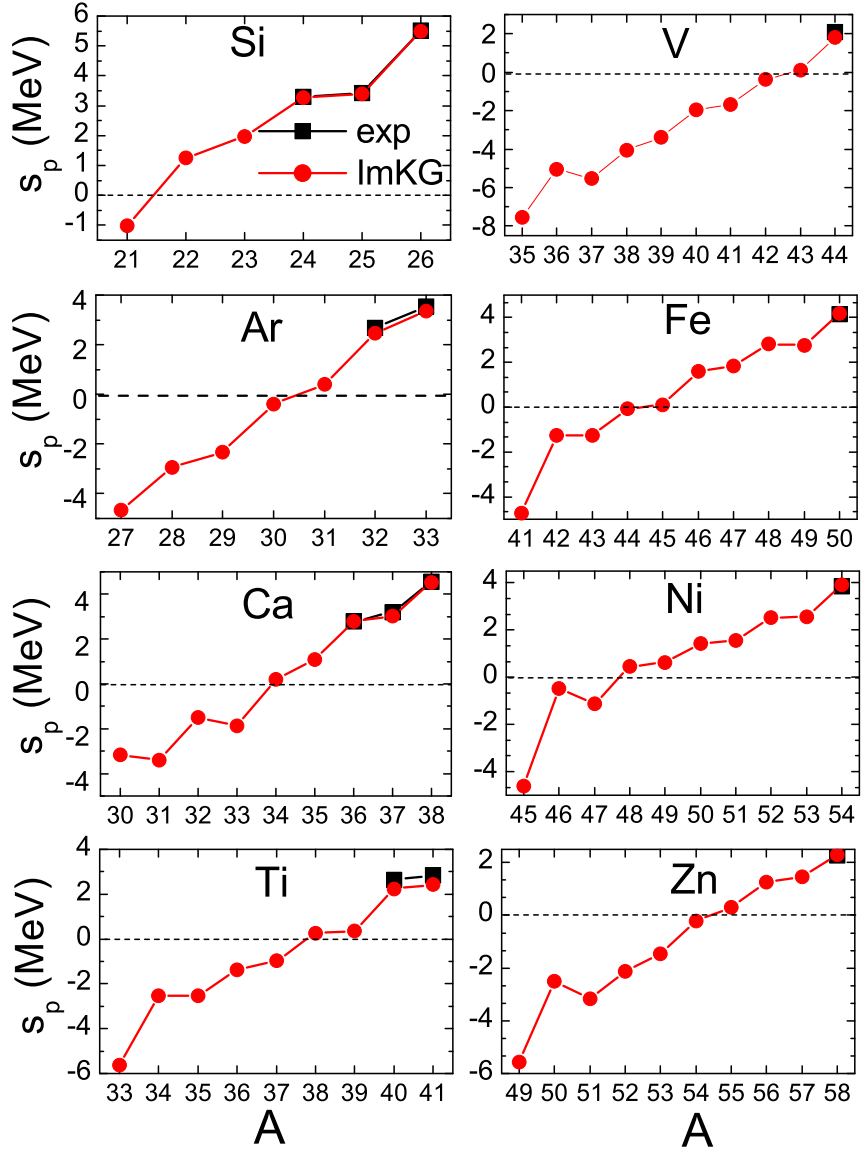}
 \caption{(Color online) Predicted one-proton separation energies (S$_{p}$ in MeV) versus mass number for proton-rich isotopes of 8 elements Si, Ar, Ca, Ti, V, Fe, Ni and Zn,
respectively.}
\end{figure}

\begin{figure}
\includegraphics[angle=-0,width= 0.7\textwidth]{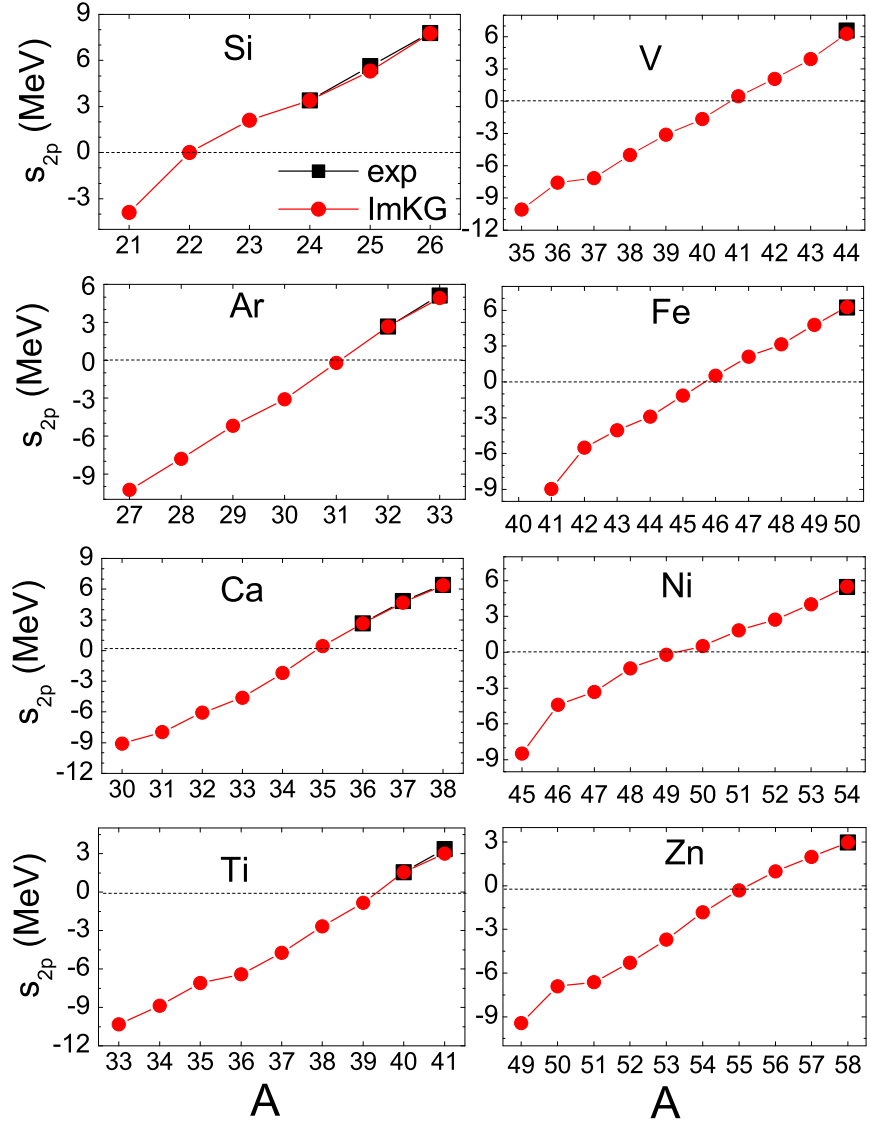}
 \caption{(Color online)Same as Fig. 5 but for two-proton separation energies S$_{2p}$.  }
\end{figure}

\begin{figure}
\includegraphics[angle=-0,width= 0.5\textwidth]{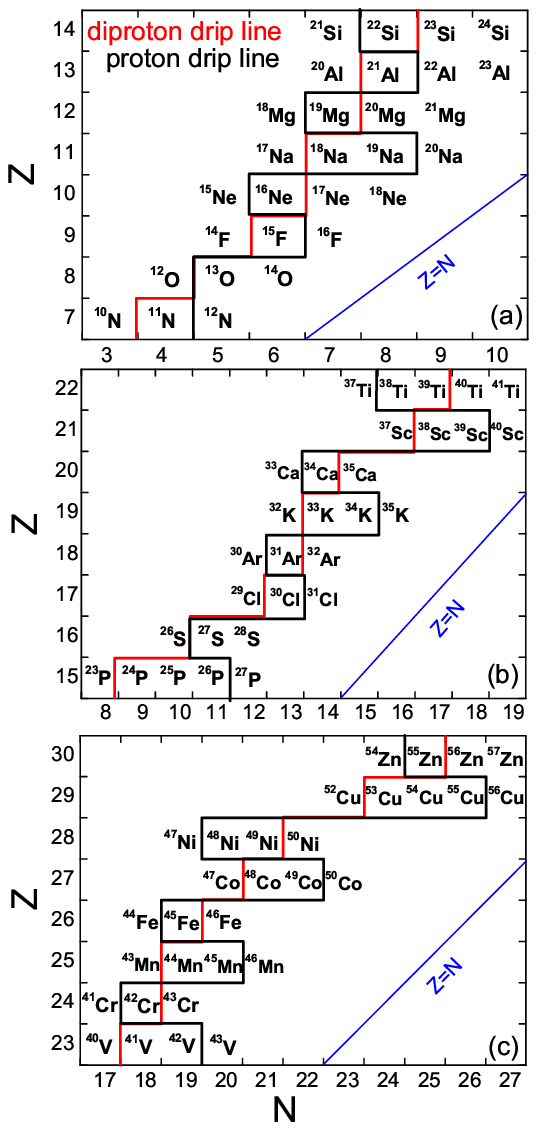}
 \caption{(Color online) Predicted positions of proton (black line) and diproton
(red line) drip lines for $Z=7-30$ with the ImKG mass relations.   }
\end{figure}

Fig. 5 and Fig. 6 show the predicted one-proton and two-proton
separation energies for  proton-rich isotopes of 8 elements Si, Ar,
Ca, Ti, V, Fe, Ni and Zn, respectively. One sees that the odd-even
staggering effect caused by the pairing interaction is evident for
one-proton separation energies, and generally disappears for the
two-proton separation energies.

The proton or diproton drip-line is the limit beyond which nuclei
become unbound, i.e., $S_{p}<0$, or $S_{2p}<0$, where $S_{p}$ and
$S_{2p}$ are the one- and two-proton separation energy. 14
proton-rich nuclei beyond the proton drip-line are observed in
experiment, they are $^{4,5}$Li, $^{7,9}$B, $^{10,11}$N ,$^{12}$O,
$^{14-16}$F, $^{18,19}$Na, $^{39}$Sc and $^{65}$As in Ref. \cite{AME2011}.
Using the separation energies of nuclei listed in Table I, the
positions of the proton and diproton drip lines are calculated for
nuclei with $Z=7-30$. The results are shown in Fig. 7. The solid
line in black denotes the proton drip line and the solid line in red is the
diproton drip line, respectively.

These nuclei with negative one-proton separation energy ($S_{p}<0$)
and positive two-proton separation energy ($S_{2p}>0$), which lie on
the left of the black line and on the right the red line, are good
candidates for one-proton emissions. We list here these nuclei
$^{11}$N, $^{15}$F, $^{18,19}$Na, $^{21}$Al, $^{24-26}$P, $^{30}$Cl,
$^{33,34}$K, $^{38,39}$Sc, $^{41,42}$V, $^{44,45}$Mn, $^{48,49}$Co
and $^{53-55}$Cu. The nuclei with positive one-proton separation
energies ($S_{p}>0$) and negative two-proton separation energies
($S_{2p}<0$) are good candidates for diproton emissions, which lie
on the left of the red line and on the right of the black line. We
list five measured proton-rich nuclei with diproton decay mode
$^{6}$Be, $^{8}$C, $^{12}$O, $^{16}$Ne and $^{19}$Mg. We also list
ten candidate nuclei with diproton decay according to the predicted
masses data in Table I, they are $^{22}$Si, $^{31}$Ar, $^{34}$Ca,
$^{38,39}$Ti, $^{42}$Cr, $^{45}$Fe, $^{48,49}$Ni and $^{55}$Zn. We
also note that if $^{A}$Z is a nucleus with possible proton
(diproton) decay mode, then the mother nucleus $^{A}$Z+$^{4}_{2}$He
is very often also a proton (diproton) decay nucleus. For example,
$^{11}$N, $^{15}$F and $^{19}$Na are good candidates for one-proton
emissions from the calculations, the three nuclei have the
relationship $^{11}$N+$^4$He $\rightarrow$ $^{15}$F, $^{15}$F+$^4$He
$\rightarrow$ $^{19}$Na.

\section{\label{sec:level2} Summary}

In summary, we propose a set of improved Kelson-Garvey mass
relations from the mass differences of mirror nuclei including
mirror nuclei far from the $\beta$-stability line as well as mirror
nuclei near the $N=Z$ line. The masses of 31 measured proton-rich
nuclei with $7\leq A\leq41$ and $-5\leq (N-Z)\leq-3$ can be
remarkably well reproduced with the proposed method. The
root-mean-square deviation is only 0.398 MeV, much smaller than the
results from the traditional Kelson-Garvey and Isobar-Mirror mass
relations. Many more masses of participating nuclei are involved in
the proposed method for the calculation of the masses of unmeasured
proton-rich nuclei. We would like to emphasize that the ImKG mass
relations proposed in this work is different from the extended
Garvey-Kelson (GK12) mass relations proposed in Ref.\cite{Barea07}.
The former is to predict the masses of the most proton-rich nuclei
by using their neutron-rich analogs mass and the masses of
participating nuclei in between, whilst the latter is to predict the
mass of a central nucleus by using the masses of 20 neighboring
nuclei around it. Binding energies and mass excesses for 144 unknown
proton-rich nuclei with $5\leq A\leq74$ are predicted systematically
by means of the ImKG mass relations. Based on the binding energies,
we have investigated the one- and two-proton separation energies,
the proton drip line and the diproton drip line in the region of
$7\leq Z\leq30$. The nuclei with positive one-proton separation
energy but negative two-proton separation energy ($S_{p}>0$,
$S_{2p}<0$) are good candidates for the study of diproton emissions.

\begin{center}
\textbf{ACKNOWLEDGEMENTS}
\end{center}

This work was supported by National Natural Science Foundation of
China, Nos. 11005003, 11275052, 10847004, 11005022 and 10975095, and innovation fund of undergraduate at Anyang Normal University (ASCX/2012-Z28).

\begin{table}
\caption{Predicted binding energy ($B_{\rm ImKG}$), mass excess
($M_{\rm ImKG}$), separation energies for one-proton (S$_{p}$) and
two-proton (S$_{2p}$) with the ImKG mass relations for 144
proton-rich nuclei (column 1) with  5$5\leq A \leq74$  are given in
columns 3, 4, 7 and 8, respectively (in MeV). The estimated
uncertainties are given in parentheses in keV. The isotopic number
$Y=N-Z$ is listed in column 2. Column 5 and 6 list the neutron-rich
analog and the measured mass excess ($M_{\rm exp}^{\rm analog}$)
with its associated error in parentheses in keV.    }

{\begin{tabular}{@{}ccccccccc@{}}\toprule

     $^{A}Z$ & $~~ Y ~~$  &   $~ $ B$_{\rm ImKG}$ & $~ $ $M_{\rm ImKG}$  &  $~ $ $^{A}Z_{\rm analog}$ & $~ $   $M_{\rm exp}^{\rm analog}$   & $~ $  $S_{p}$  & $~ $  $S_{2p}$ \\
             &   &   $~ $  MeV (keV)   &  $~ $   MeV (keV)  &  $~ $                & $~ $       MeV (keV)              & $~ $  MeV (keV)    & $~ $  MeV (keV)    \\
\hline
$^{52}$Co  &       -2     &         432.983(15)  &    -34.396(15)   &    $^{52}$Mn    &       -50.707(2)      &    1.466(21)      &    6.348(15)    \\
$^{56}$Cu  &       -2     &         453.218(1)   &   -38.571(1)     &    $^{56}$Co    &       -56.040(1)      &    0.526(1)       &    5.141(1)     \\
$^{60}$Ga  &       -2     &         499.975(55)  &   -39.947(55)    &    $^{60}$Cu    &       -58.345(2)      &    0.023(66)        &    2.860(55)    \\
$^{62}$Ge  &       -2     &         517.675(66)  &    -42.287(66)   &    $^{62}$Zn    &       -61.167(1)      &    2.489(107)     &    2.693(66)    \\
$^{64}$As  &       -2     &         530.412(92)  &   -39.664(92)    &    $^{64}$Ga    &       -58.834(1)      &    0.034(114)       &    2.258(92)    \\
$^{66}$Se  &       -2     &         548.166(108) &   -42.058(108)   &    $^{66}$Ge    &       -61.607(2)      &    2.412(128)       &    2.322(108)   \\
$^{68}$Br  &       -2     &         560.507(125) &    -39.038(125)  &    $^{68}$As    &       -58.895(2)      &   -0.251(179)      &    1.589(125)   \\
$^{70}$Kr  &       -2     &         578.734(166) &    -41.904(166)  &    $^{70}$Se    &       -61.930(2)      &    2.722(181)      &    2.294(166)   \\
$^{72}$Rb  &       -2     &         590.613(163) &    -38.422(163)  &    $^{72}$Br    &       -59.067(7)      &   -0.614(163)      &    1.577(163)   \\
$^{74}$Sr  &       -2     &         608.812(242) &    -41.262(242)  &    $^{74}$Kr    &       -62.332(2)      &    2.474(242)     &    1.901(242)   \\

$^{ 5}$Be  &       -3     &          3.129(120)  &     34.097(120)   &    $^{ 5}$H     &        32.892(89)    &    -1.759(165)    &    -4.589(120)    \\
$^{43}$V   &       -3     &         347.063(42)  &    -17.990(42)    &    $^{43}$Ca    &      -38.409(0.2)       &     0.094(43)     &    3.926(42)    \\
$^{45}$Cr  &       -3     &         364.134(42)  &    -19.700(42)    &    $^{45}$Sc     &     -41.070(0.6)    &     3.168(47)     &    4.958(43)    \\
$^{47}$Mn  &       -3     &         382.415(43)  &    -22.621(43)    &    $^{47}$Ti    &      -44.936(0.4)    &     0.358(48)      &    5.320(46)    \\
$^{49}$Fe  &       -3     &          399.921(45)  &    -24.767(45)   &    $^{49}$V     &      -47.961(0.9)    &     2.744(53)     &    4.787(47)    \\
$^{51}$Co  &       -3     &         417.927(27)  &    -27.412(27)    &    $^{51}$Cr    &      -51.451(0.9)     &     0.179(39)     &    4.376(36)    \\
$^{53}$Ni  &       -3     &         435.544(20)  &    -29.668(20)    &    $^{53}$Mn    &      -54.689(0.6)       &     2.561(25)     &    4.026(25)    \\
$^{55}$Cu  &       -3     &         452.943(16)  &    -31.707(16)    &    $^{55}$Fe    &      -57.481(0.5)    &    -0.276(17)     &    3.628(17)    \\
$^{57}$Zn  &       -3     &         469.325(16)  &    -32.730(16)    &    $^{57}$Co    &      -59.345(0.5)    &     1.447(17)      &    1.973(17)    \\

$^{ 6}$B   &       -4     &           0.360(264) &     44.235(264)   &    $^{ 6}$H     &       41.876(254)   &    -2.769(290)    &   -4.528(288)    \\
$^{22}$Al  &       -4     &         149.429(12)  &     17.969(12)    &    $^{22}$F     &        2.793(12)      &     0.149(13)     &    3.378(13)    \\
$^{26}$P   &       -4     &         186.917(5)   &     11.203(5)     &    $^{26}$Na     &      -6.861(3.5)     &    -0.119(6)      &    3.291(5)    \\
$^{30}$Cl  &       -4     &         224.010(17)  &      4.830(17)    &    $^{30}$Al     &     -15.872(14)    &    -0.611(17)      &    2.653(17)    \\

\botrule
\end{tabular}}
\end{table}

\begin{table}
{\begin{tabular}{@{}ccccccccc@{}}\toprule

     $^{A}Z$ & $~~ Y ~~$  &   $~ $ B$_{\rm ImKG}$ & $~ $ $M_{\rm ImKG}$  &  $~ $ $^{A}Z_{\rm analog}$ & $~ $   $M_{\rm exp}^{\rm analog}$   & $~ $  $S_{p}$  & $~ $  $S_{2p}$ \\
\hline
$^{34}$K   &       -4     &         261.023(1)   &         -1.462(1)     &    $^{34}$P      &     -24.549(0.8)   &    -0.642(2)      &    2.715(2)    \\
$^{38}$Sc  &       -4     &         294.969(4)   &      -4.687(4)     &    $^{38}$Cl     &     -29.798(0.1)   &    -1.155(4)         &    1.857(4)    \\
$^{42}$V   &       -4    &         329.011(63)  &     -8.009(63)    &    $^{42}$K      &      -35.022(0.1)   &    -0.351(64)     &    2.071(63)    \\
$^{44}$Cr  &       -4     &         349.911(22)  &    -13.548(22)    &    $^{44}$Ca     &      -41.469(0.3)   &     2.848(48)      &    2.942(23)    \\
$^{46}$Mn  &       -4     &         364.406(50)  &    -12.683(50)    &    $^{46}$Sc     &      -41.760(0.6)  &     0.272(66)       &    3.440(54)    \\
$^{48}$Fe  &       -4    &         385.225(29)  &    -18.142(29)    &    $^{48}$Ti     &      -48.492(0.4)   &     2.810(52)       &    3.168(37)    \\
$^{50}$Co  &       -4      &         400.188(36)  &    -17.745(36)    &    $^{50}$V      &      -49.224(0.9)   &     0.267(58)     &    3.011(46)    \\
$^{52}$Ni  &       -4     &         420.472(27)  &    -22.668(27)    &    $^{52}$Cr     &      -55.418(0.6)   &     2.545(38)      &    2.725(39)    \\
$^{54}$Cu  &       -4      &         434.930(12)  &    -21.765(12)    &    $^{54}$Mn     &      -55.557(1.2)   &    -0.614(24)     &    1.947(20)    \\
$^{56}$Zn  &       -4     &         454.211(20)  &    -25.686(20)    &    $^{56}$Fe     &      -60.606(0.5)  &     1.268(27)       &    0.992(21)    \\

$^{13}$F   &       -5     &         53.666(16)   &   43.386(16)    &    $^{13}$Be    &      33.208(10)       &    -3.235(18)       &   -3.425(16)    \\
$^{15}$Ne  &       -5     &         71.613(23)   &   41.555(23)    &    $^{15}$B     &      28.964(22)       &    -1.270(33)       &   -3.532(23)    \\
$^{17}$Na  &       -5     &         93.481(23)   &   35.346(23)    &    $^{17}$C     &      21.032(17)       &    -3.561(26)       &   -3.328(23)    \\
$^{21}$Al  &       -5     &        133.304(12)   &  26.024(12)     &    $^{21}$O     &      8.062(12)        &    -1.265(13)       &    1.496(13)    \\
$^{23}$Si  &       -5     &        151.394(100)  &  23.294(100)    &    $^{23}$F     &      3.310(100)       &    1.965(101)       &    2.114(100)    \\
$^{25}$P   &       -5     &        170.495(44)   &   19.553(44)     &    $^{25}$Ne    &      -2.060(45)      &    -1.507(45)       &    1.770(45)    \\
$^{27}$S   &       -5    &        187.474(13)   &   17.934(13)     &    $^{27}$Na    &      -5.518(4)        &     0.557(14)       &    0.438(13)    \\
$^{29}$Cl  &       -5        &        206.427(17)   &   14.341(17)     &    $^{29}$Mg    &      -10.603(11)      &    -2.682(20)       &   -0.341(18)    \\
$^{31}$Ar  &       -5        &        224.410(21)   &   11.719(21)     &    $^{31}$Al    &      -14.955(20)      &     0.400(28)       &   -0.212(22)    \\
$^{33}$K   &       -5    &        243.971(6)    &    7.518(6)      &    $^{33}$Si    &      -20.514(0.7)     &    -2.414(10)       &    0.040(7)    \\
$^{35}$Ca  &       -5    &        262.100(3)    &    4.750(3)      &    $^{35}$P     &      -24.858(2)       &     1.077(6)      &    0.435(4)    \\
$^{37}$Sc  &       -5    &        278.413(4)    &    3.797(4)      &    $^{37}$S     &      -26.896(0.2)       &    -2.961(4)        &   -0.370(4)    \\
$^{39}$Ti  &       -5      &        295.276(49)   &      2.294(49)     &    $^{39}$Cl    &      -29.800(1.7)       &     0.307(50)     &   -0.848(50)    \\
$^{41}$V   &       -5    &        312.882(43)   &    0.048(43)     &    $^{41}$Ar    &      -33.068(0.4)       &    -1.679(68)     &    0.447(44)    \\
$^{43}$Cr    &       -5        &        330.352(52)   &   -2.061(52)     &    $^{43}$K     &      -36.575(0.4)     &     1.341(82)     &    0.990(53)    \\
$^{45}$Mn    &       -5        &        348.837(38)   &   -5.185(38)     &    $^{45}$Ca    &      -40.812(0.4)     &    -1.074(44)     &    1.773(57)    \\
$^{47}$Fe    &       -5        &        366.237(42)   &   -7.226(42)     &    $^{47}$Sc    &      -44.337(2)         &     1.832(66)       &    2.103(60)    \\
$^{49}$Co    &       -5        &        384.300(23)   &   -9.928(23)     &    $^{49}$Ti    &      -48.563(0.4)     &    -0.925(38)     &    1.885(49)    \\
$^{51}$Ni  &       -5      &        401.747(36)   &  -12.014(36)     &    $^{51}$V     &      -52.204(0.9)     &     1.558(51)     &    1.826(58)    \\

\botrule
\end{tabular}}
\end{table}

\begin{table}
{\begin{tabular}{@{}ccccccccc@{}}\toprule

     $^{A}Z$  & $~~ Y ~~ $   &    B$_{\rm ImKG}$ & $~ $ $M_{\rm ImKG}$  &  $~ $ $^{A}Z_{\rm analog}$ & $~ $  $M_{\rm exp}^{\rm analog}$ &  $~ $  $S_{p}$  & $~ $  $S_{2p}$ \\
\hline
$^{53}$Cu    &       -5        &        418.858(23)   &  -13.766(23)     &    $^{53}$Cr    &      -55.286(0.6)     &    -1.614(35)      &   0.932(35)  \\
$^{55}$Zn    &       -5        &        435.227(18)   &  -14.774(18)     &    $^{55}$Mn    &      -57.712(0.4)     &     0.297(22)    &  -0.317(27)    \\

$^{16}$Na  &       -6     &         67.264(30)   &    53.272(30)    &    $^{16}$B     &       37.120(27)      &    -4.427(38)      &    -5.697(38)    \\
$^{18}$Mg  &       -6     &         92.589(32)   &    43.306(32)    &    $^{18}$C     &       24.920(30)      &    -0.672(39)      &    -4.233(34)    \\
$^{20}$Al  &       -6     &        108.982(56)   &    42.274(56)    &    $^{20}$N     &       21.765(56)      &    -2.890(60)      &    -2.403(60)    \\
$^{22}$Si  &       -6     &        134.565(57)   &    32.051(57)    &    $^{22}$O     &       9.282(57)       &     1.261(59)      &    -0.004(57)    \\
$^{24}$P   &       -6     &        149.528(72)   &    32.448(72)    &    $^{24}$F     &       7.560(72)       &    -1.866(124)     &     0.099(74)    \\
$^{26}$S   &       -6     &        169.856(21)   &    27.481(21)    &    $^{26}$Ne    &       0.479(18)       &    -0.639(50)      &    -2.146(21)    \\
$^{28}$Cl  &       -6     &        184.650(12)   &    28.047(12)    &    $^{28}$Na    &      -0.988(10)       &    -2.824(18)      &    -2.267(14)    \\
$^{30}$Ar  &       -6     &        206.034(17)   &    22.024(17)    &    $^{30}$Mg    &      -8.892(13)       &    -0.394(25)      &    -3.075(20)    \\
$^{32}$K   &       -6     &        221.665(86)   &    21.753(86)    &    $^{32}$Al    &     -11.062(86)       &    -2.745(89)      &    -2.345(88)    \\
$^{34}$Ca  &       -6     &        244.182(16)   &    14.596(16)    &    $^{34}$Si    &     -19.957(14)       &     0.211(17)    &    -2.203(17)    \\
$^{36}$Sc  &       -6     &        258.300(13)   &    15.838(13)    &    $^{36}$P     &     -20.251(13)       &    -3.800(14)    &    -2.722(13)    \\
$^{38}$Ti  &       -6     &        278.683(27)   &    10.816(27)    &    $^{38}$S     &     -26.861(7)        &     0.270(27)    &    -2.691(27)    \\
$^{40}$V   &       -6     &        293.334(52)   &    11.525(52)    &    $^{40}$Cl    &     -27.558(32)       &    -1.942(72)      &    -1.635(52)    \\
$^{42}$Cr  &       -6     &        314.025(32)   &     6.194(32)    &    $^{42}$Ar    &     -34.423(6)        &     1.143(54)      &    -0.536(62)    \\
$^{44}$Mn  &       -6     &        329.140(59)   &     6.439(59)    &    $^{44}$K     &     -35.782(0.4)      &    -1.212(79)      &    0.129(87)    \\
$^{46}$Fe  &       -6     &        350.419(30)   &     0.522(30)    &    $^{46}$Ca    &     -43.140(2)        &     1.582(48)      &    0.508(37)    \\
$^{48}$Co  &       -6     &        365.407(38)   &     0.893(38)    &    $^{48}$Sc    &     -44.503(5)        &    -0.830(57)      &    1.002(63)    \\
$^{50}$Ni  &       -6     &        385.746(26)   &    -4.085(26)    &    $^{50}$Ti    &     -51.431(0.4)      &     1.446(35)      &    0.522(39)    \\
$^{52}$Cu  &       -6     &        399.506(33)   &    -2.484(33)    &    $^{52}$V     &     -51.444(0.9)      &    -2.241(49)      &  -0.683(49)       \\
$^{54}$Zn  &       -6     &        418.649(25)   &    -6.267(25)    &    $^{54}$Cr    &     -56.934(0.6)      &    -0.210(34)      &  -1.824(37)       \\

$^{19}$Al  &       -7    &       88.100(99)   &   55.085(99)    &    $^{19}$C     &      32.414(98)          &   -4.489(104)   &    -5.161(102)    \\
$^{21}$Si  &       -7    &      107.977(96)   &     50.568(96)    &    $^{21}$N     &      25.251(95)          &   -1.005(111)   &    -3.895(98)    \\
$^{23}$P   &       -7    &      131.610(90)   &     42.295(90)    &    $^{23}$O     &      14.621(90)          &   -2.955(107)   &    -1.694(91)    \\
$^{25}$S   &       -7    &      147.625(76)   &     41.641(76)    &    $^{25}$F     &      11.364(75)          &   -1.904(106)   &    -3.769(126)    \\
$^{27}$Cl  &       -7    &      165.005(66)   &     39.620(66)    &    $^{27}$Ne    &      7.036(65)           &   -4.851(69)   &    -5.490(80)    \\
$^{29}$Ar  &       -7    &      182.309(14)   &     37.677(14)    &    $^{29}$Na    &      2.670(12)           &   -2.341(19)   &    -5.165(19)    \\
$^{31}$K   &       -7    &      201.481(20)   &     33.866(20)    &    $^{31}$Mg    &      -3.190(17)          &   -4.554(27)   &    -4.947(27)    \\

\botrule
\end{tabular}}
\end{table}

\begin{table}
{\begin{tabular}{@{}ccccccccc@{}}\toprule

     $^{A}Z$ & $~~ Y ~~$  &   $~ $ B$_{\rm ImKG}$ & $~ $ $M_{\rm ImKG}$  &  $~ $ $^{A}Z_{\rm analog}$ & $~ $   $M_{\rm exp}^{\rm analog}$   & $~ $  $S_{p}$  & $~ $  $S_{2p}$ \\
\hline

$^{33}$Ca  &       -7    &      219.808(68)   &     30.898(68)    &    $^{33}$Al    &      -8.437(68)          &   -1.857(110)   &    -4.602(72)    \\
$^{35}$Sc  &       -7    &      239.393(39)   &     26.674(39)    &    $^{35}$Si    &      -14.360(38)         &   -4.789(42)   &    -4.579(40)    \\
$^{37}$Ti  &       -7    &      257.350(61)   &     24.077(61)    &    $^{37}$P     &      -18.996(38)         &   -0.950(62)   &    -4.750(61)    \\
$^{39}$V   &       -7    &      275.285(64)   &     21.503(64)    &    $^{39}$S     &      -23.162(50)         &   -3.398(70)   &    -3.129(64)    \\
$^{41}$Cr  &       -7    &      292.871(75)   &     19.277(75)    &    $^{41}$Cl    &      -27.307(68)         &   -0.463(91)   &    -2.405(90)    \\
$^{43}$Mn  &       -7     &      311.195(44)   &   16.314(44)    &    $^{43}$Ar    &      -32.010(5)           &   -2.831(55)   &    -1.688(62)    \\
$^{45}$Fe  &       -7     &      329.229(55)   &   13.640(55)    &    $^{45}$K     &      -36.616(0.5)         &    0.089(81)   &    -1.123(76)    \\
$^{47}$Co  &       -7     &      348.628(26)      &    9.601(26)    &    $^{47}$Ca    &      -42.345(2)          &   -1.791(40)   &    -0.209(46)    \\
$^{49}$Ni  &       -7     &      366.043(39)      &    7.547(39)    &    $^{49}$Sc    &      -46.560(3)          &    0.636(55)   &    -0.194(58)    \\
$^{51}$Cu  &       -7     &      382.566(22)      &    6.384(22)    &    $^{51}$Ti    &      -49.732(0.6)        &   -3.180(34)   &    -1.734(33)    \\
$^{53}$Zn  &       -7     &      398.054(35)      &    6.257(35)    &    $^{53}$V     &      -51.850(3)          &   -1.452(48)    &   -3.693(50)    \\

$^{22}$P   &       -8      &     103.574(192)   &  62.260(192)    &    $^{22}$N     &      32.039(192)         &    -4.403(215)    &   -5.408(201)    \\
$^{24}$S   &       -8    &     129.631(111)   &  51.563(111)    &    $^{24}$O     &      18.500(110)         &    -1.980(143)      &   -4.934(125)    \\
$^{26}$Cl  &       -8      &     142.028(78)    &  54.526(78)     &    $^{26}$F     &      18.665(77)          &    -5.596(109)    &   -7.499(107)    \\
$^{28}$Ar  &       -8      &     162.067(96)    &  49.848(96)     &    $^{28}$Ne    &      11.292(96)          &    -2.939(116)    &   -7.790(98)    \\
$^{30}$K   &       -8     &     177.753(24)    &  49.522(24)     &    $^{30}$Na    &      8.374(23)           &    -4.556(28)      &   -6.897(28)    \\
$^{32}$Ca  &       -8     &     199.970(21)    &  42.666(21)     &    $^{32}$Mg    &      -0.912(18)          &    -1.511(30)      &   -6.064(28)    \\
$^{34}$Sc  &       -8     &     215.226(60)    &  42.770(60)     &    $^{34}$Al    &      -3.047(60)          &    -4.583(92)      &   -6.440(105)    \\
$^{36}$Ti  &       -8     &     237.755(74)    &  35.601(74)     &    $^{36}$Si    &      -12.418(57)         &    -1.638(84)      &   -6.427(76)    \\
$^{38}$V   &       -8     &     253.298(81)    &  35.419(81)     &    $^{38}$P     &      -14.643(71)         &    -4.053(102)    &   -5.003(83)    \\
$^{40}$Cr  &       -8     &     274.841(118)   &  29.236(118)    &    $^{40}$S     &      -22.930(114)        &    -0.444(134)     &   -3.843(121)    \\
$^{42}$Mn  &       -8     &     290.059(150)   &  29.378(150)    &    $^{42}$Cl    &      -24.913(144)        &    -2.812(168)     &   -3.275(159)    \\
$^{44}$Fe    &       -8     &     311.112(39)    &  23.685(39)     &    $^{44}$Ar    &      -32.673(2)          &    -0.082(59)    &   -2.913(51)    \\
$^{46}$Co    &       -8     &     327.032(53)    &  23.127(53)     &    $^{46}$K     &      -35.414(0.7)        &    -2.198(76)    &   -2.109(79)    \\
$^{48}$Ni    &       -8     &     349.085(29)    &  16.433(29)     &    $^{48}$Ca    &      -44.224(2)          &     0.457(39)    &   -1.334(42)    \\
$^{50}$Cu  &       -8     &     362.548(40)    &  18.331(40)     &    $^{50}$Sc    &      -44.546(15)         &    -3.495(56)      &   -2.859(56)    \\
$^{52}$Zn    &       -8     &     380.445(27)    &  15.794(27)     &    $^{52}$Ti    &      -49.469(7)          &    -2.121(35)    &   -5.301(37)      \\

$^{25}$Cl  &       -9    &       122.489(111)   &   65.994(111)  &    $^{25}$O     &     27.348(111)             &   -7.142(157)       &  -9.121(143 )    \\
$^{27}$Ar  &       -9    &       137.381(190)   &   66.462(190)  &    $^{27}$F     &     24.630(190)           &   -4.647(206)     &  -10.244(205)    \\

\botrule
\end{tabular}}
\end{table}

\begin{table}
{\begin{tabular}{@{}ccccccccc@{}}\toprule

     $^{A}Z$ & $~~ Y ~~$  &   $~ $ B$_{\rm ImKG}$ & $~ $ $M_{\rm ImKG}$  &  $~ $ $^{A}Z_{\rm analog}$ & $~ $   $M_{\rm exp}^{\rm analog}$   & $~ $  $S_{p}$  & $~ $  $S_{2p}$ \\
\hline

$^{29}$K   &       -9      &     156.107(100)   &   63.097(100)  &    $^{29}$Ne    &     18.400(100)           &   -5.960(139)     &  -8.899(120)    \\
$^{31}$Ca  &       -9    &     174.356(103)   &     60.208(103)  &    $^{31}$Na    &     12.540(103)             &   -3.397(106)       &  -7.954(104)    \\
$^{33}$Sc  &       -9    &     194.918(23)    &   55.006(23)   &    $^{33}$Mg    &      4.947(22)            &   -5.052(31)      &  -6.562(31)    \\
$^{35}$Ti  &       -9     &     212.704(70)    &    52.581(70)   &    $^{35}$Al    &     -0.220(70)              &   -2.522(93)      &  -7.105(98)    \\
$^{37}$V   &       -9     &     232.233(92)    &    48.413(92)   &    $^{37}$Si    &     -6.594(83)              &   -5.523(118)     &  -7.161(100)    \\
$^{39}$Cr  &       -9     &      251.238(87)    &   44.768(87)   &    $^{39}$P     &     -12.795(81)             &   -2.060(119)       &  -6.112(106)    \\
$^{41}$Mn  &       -9     &      270.581(74)    &   40.785(74)   &    $^{41}$S     &     -19.089(61)             &   -4.260(139)       &  -4.704(98)    \\
$^{43}$Fe  &       -9     &      288.809(209)   &   37.917(209)  &    $^{43}$Cl    &     -24.408(206)            &   -1.250(257)       &  -4.062(222)    \\
$^{45}$Co  &       -9     &      307.214(34)    &   34.873(34)   &    $^{45}$Ar    &     -29.771(0.5)            &   -3.899(52)      &  -3.981(56)    \\
$^{47}$Ni  &       -9     &      325.905(54)    &   31.542(54)   &    $^{47}$K     &     -35.709(2.5)            &   -1.126(75)      &  -3.324(77)    \\
$^{49}$Cu  &       -9     &      344.679(26)    &   28.129(26)   &    $^{49}$Ca    &     -41.299(2)            &   -4.406(39)        &  -3.949(37)    \\
$^{51}$Zn  &       -9     &      359.371(44)    &   28.796(44)   &    $^{51}$Sc    &     -43.228(20)           &   -3.177(60)        &  -6.672(59)    \\

$^{28}$K     &      -10    &    130.729(503)   &    80.403(503)  &    $^{28}$F     &      33.115(503)          &   -6.652(538)     &    -11.300(509)    \\
$^{30}$Ca  &       -10    &    152.958(281)   &    73.535(281)  &    $^{30}$Ne    &      23.040(280)          &   -3.149(298)      &     -9.110(297)    \\
$^{32}$Sc  &       -10    &    169.651(120)   &    72.202(120)  &    $^{32}$Na    &      18.810(120)          &   -4.705(159)      &     -8.102(123)    \\
$^{34}$Ti  &       -10    &    192.396(102)   &    64.817(102)  &    $^{34}$Mg    &       8.560(90)           &   -2.522(105)      &     -7.574(105)    \\
$^{36}$V   &       -10    &    207.654(107)   &    64.920(107)  &    $^{36}$Al    &       5.950(100)          &   -5.050(129)     &     -7.572(123)    \\
$^{38}$Cr  &       -10    &    230.442(76)    &    57.492(76)   &    $^{38}$Si    &      -4.170(70)           &   -1.790(120)      &     -7.313(106)    \\
$^{40}$Mn  &       -10    &    246.974(119)   &    56.321(119)  &    $^{40}$P     &      -8.074(111)          &   -4.264(147)      &     -6.324(145)    \\
$^{42}$Fe    &       -10    &    269.335(129)   &    49.320(129)  &    $^{42}$S     &     -17.678(124)          &   -1.246(150)      &     -5.506(175)    \\
$^{44}$Co    &       -10    &    284.960(142)   &    49.055(142)  &    $^{44}$Cl    &     -20.605(137)          &   -3.849(253)      &     -5.099(207)    \\
$^{46}$Ni    &       -10    &    306.737(56)    &    42.638(56)   &    $^{46}$Ar    &     -29.729(41)           &   -0.477(65)       &     -4.376(68)    \\
$^{48}$Cu    &       -10    &    322.026(53)    &    42.711(53)   &    $^{48}$K     &     -32.285(2)            &   -3.880(75)       &     -5.006(74)    \\
$^{50}$Zn  &       -10    &    342.168(29)    &    37.928(29)   &    $^{50}$Ca    &     -39.588(3)            &   -2.510(39)         &     -6.916(41)    \\

$^{31}$Sc    &       -11    &     145.263(1620)  &  88.519(1620) &    $^{31}$Ne    &      30.820(1620)       &    -7.695(1644)   &   -10.844(1623)    \\
$^{33}$Ti    &       -11    &     164.042(598)   &  85.100(598)  &    $^{33}$Na    &       23.967(596)       &    -5.609(610)      &   -10.314(607)    \\
$^{35}$V     &       -11    &     184.833(185)   &  79.670(185)  &    $^{35}$Mg    &       15.640(180)       &    -7.564(211)      &   -10.085(186)    \\
$^{37}$Cr    &       -11    &     202.764(124)   &  77.097(124)  &    $^{37}$Al    &       9.810(120)          &    -4.890(164)    &   -9.940(143)    \\
$^{39}$Mn    &       -11    &     222.689(100)   &  72.534(100)  &    $^{39}$Si    &       2.320(90)           &    -7.753(126)    &   -9.543(136)    \\

\botrule
\end{tabular}}
\end{table}

\begin{table}
{\begin{tabular}{@{}ccccccccc@{}}\toprule

     $^{A}Z$ & $~~ Y ~~$  &   $~ $ B$_{\rm ImKG}$ & $~ $ $M_{\rm ImKG}$  &  $~ $ $^{A}Z_{\rm analog}$ & $~ $   $M_{\rm exp}^{\rm analog}$   & $~ $  $S_{p}$  & $~ $  $S_{2p}$ \\
 \hline
$^{41}$Fe    &       -11    &     242.261(89)    &  68.322(89)   &    $^{41}$P     &      -4.980(80)           &    -4.713(149)    &   -8.977(124)    \\
$^{43}$Co    &       -11    &     261.823(105)   &   64.121(105)  &    $^{43}$S     &     -12.070(100)         &    -7.512(167)    &   -8.758(129)    \\
$^{45}$Ni    &       -11    &     280.345(106)   &   60.960(106)  &    $^{45}$Cl    &     -18.360(100)         &    -4.615(177)    &   -8.464(235)    \\
$^{47}$Cu    &       -11    &     299.753(97)    &   56.912(97)   &    $^{47}$Ar    &     -25.210(90)           &    -6.985(112)    &   -7.462(103)    \\
$^{49}$Zn    &       -11    &     316.471(54)    &   55.554(54)   &    $^{49}$K     &     -29.612(3)          &    -5.555(75)       &   -9.434(76)    \\

\botrule
\end{tabular}}
\end{table}

\end{document}